\documentclass[12pt]{iopart}

 \usepackage{iopams}

\newtheorem{theorem}{Theorem}
\newtheorem{prop}[theorem]{Proposition}
\newtheorem{corollary}[theorem]{Corollary}

\newcommand{\pr}[1]{${\rm #1}'$}

\begin{document}

\title{Is the Lorentz contraction inevitable in the special theory of relativity?}

\author{Sadanand D Agashe}

\address{Department of Electrical Engineering, Indian Institute of Technology Bombay, Powai, Mumbai-76, India-400076}
\ead{eesdaia@ee.iitb.ac.in}
\begin{abstract}
We look afresh at the deduction of the ``Lorentz contraction'' of a ``rod'' from the Lorentz transformation equations of the special theory of relativity.  We show that under special conditions, which include acceleration of the ``rod'',  length ``expansion'' and ``invariance'' are possible, and thus these are purely kinematical phenomena. We then take a closer look at Einstein's two famous papers on the special theory of relativity and point out a flaw in his argument. It seems that it is possible to have the times of the clocks of two, and indeed all, inertial systems to agree with each other.

\end{abstract}

\maketitle

\section{Usual scenario for the Lorentz contraction}

Most authors who deduce the Lorentz contraction from the Lorentz transformation consider the following scenario. Let $S$ denote the reference frame with origin $O$ and with coordinates $x$, $y$, $z$, and time $t$. Let $S'$ be another reference frame whose origin $O'$ moves with \emph{speed} $v$ ($v > 0$), relative to $S$, in the direction of the positive $x$-axis of $S$, and let $x'$, $y'$, $z'$ denote the coordinates in $S'$, and $t'$ the time in $S'$. Let us suppose also that the origins $O$ and $O'$ of $S$ and $S'$ coincide at time $t=0$ which is also the time $t'=0$. The $y'$- and $z'$- axes of $S'$ are assumed to remain parallel to the $y$- and $z$- axes of $S$, respectively. Then, the coordinates $x$, $y$, $z$, and time $t$ in $S$ of an event are related to the coordinates $x'$, $y'$, $z'$, and time $t'$ in $S'$ of the same event by the following Lorentz transformation:
\begin{equation}
\label{LT}
\eqalign
x' = \beta (x - vt), \\
y' = y, \\
z' = z, \\
t' = \beta (t-\frac{vx}{c^2})
\end{equation}
where $c$ denotes the speed of light in $S$ \emph{and} $S'$ and $\beta = 1/ \sqrt{(1-v^2/c^2)}$. Note that $\beta > 1$.  

The transformation (\ref{LT}) has an inverse, namely: 
\begin{equation}
\label{ILT}
\eqalign
x = \beta (x' + vt'), \\
y = y', \\
z = z', \\
t = \beta (t'+\frac{vx'}{c^2}).
\end{equation}
If the origin $O'$ of $S'$ were to move in the direction of the negative $x$- axis of $S$, with \emph{speed} $v$ ($v > 0$) then in (\ref{LT}) and (\ref{ILT}), we would have to replace $v$ by $-v$. 

\section{Deduction of the Lorentz contraction}
In deducing the Lorentz contraction in such a scenario, most authors talk about a rigid rod lying \emph{at rest} on the $x'$- axis of the moving system $S'$. The ends $P1$ and $P2$ of this rod can thus be thought of as a \emph{series} of events: $P1\equiv \{ (x_1',0,0,t')\}$ and $P2\equiv \{ (x_2',0,0,\bar{t}')\}$, with $x_2' - x_1' = l > 0$, say, so that we can call $l$ the constant (in $S'$) length of the rod, and so we are justified in calling the rod ``rigid'' in $S'$. Next, one shows that although the rod is at rest in $S'$, it is ``observed'' to be moving in $S$ with speed $v$, of course. Further, as it moves in $S$, its length remains constant in $S$, and so it is rigid in $S$ also. However, its length in $S$ is different from its length $l$ in $S'$, and is, infact,  $\frac{1}{\beta} l$, which is smaller than $l$. Hence the term ``contraction''. Indeed, using (\ref{ILT}), at any time $t$ of $S$, the coordinates of $P1$ in $S$ are $(\beta(x_1'+vt_1'),0,0)$, and the coordinates of $P2$ in $S$ are $(\beta(x_2'+vt_2'),0,0)$, where $t_1'$ and $t_2'$ in $S'$ correspond to a common time $t$ in $S$ and so, 
\begin{eqnarray}
\label{Time}
t &= \beta(t_1' + \frac{vx_1'}{c^2}) = \beta(t_2' + \frac{vx_2'}{c^2}).
\end{eqnarray}
The distance between $P1$ and $P2$ in $S$ at time $t$, and, thus, the length of the rod in $S$ at time $t$ are given by 
\begin{eqnarray}
\beta(x_2' + vt_2')-\beta(x_1' + vt_1') &= \beta(x_2' - x_1') + \beta v (t_2'-t_1') \\
&= \beta(x_2' - x_1') - \beta v  \frac{v}{c^2} (x_2'-x_1') \\
&= \beta(1 - \frac{v^2}{c^2}) (x_2'-x_1') \\
&= \frac{1}{\beta} l.
\end{eqnarray}
All this is very familiar and is written only to fix the notation and to avoid misunderstanding. Note that one could allow the rod to be anywhere in the space of $S'$, provided it is parallel to the $x'$-axis. 

\section{What happens to a rod moving with an arbitrary velocity?}
Now, this business of considering the rod \emph{at rest} in a \emph{moving} frame of reference goes back to the early days of the special theory of relativity. One could have talked about a rod lying at rest in the \emph{first-mentioned} frame, namely, $S$, and then considered its history as observed from the \emph{second-mentioned} frame, namely, $S'$, assumed to be moving uniformly relative to $S$. (This is, indeed, pointed out by many authors.) Leaving that aside, no authors seem to have considered a rod \emph{rigidly moving in $S$} and its history in $S'$.  This is what we will do, deriving some surprising consequences. (In one excellent textbook \cite{Rosser1964}, in Exercise 3.11, p.135, the possibility of the times $t$ and $t'$ of a moving point being equal is explored.) 

So, let a point $P1$ have a history- or motion! - in $S$, given by the series of events $\{ (x_0 + ut, 0,0,t) \}$, thus $P1$ moves uniformly in $S$ with speed $|u|$ in the direction of the positive $x$- axis of $S$ if $u > 0$ and in the opposite direction if $u < 0$. Let another point have the motion $\{ (x_0 + l + u\bar{t}, 0,0,\bar{t}) \}$, with $l > 0$. Thus $P2$ also moves in $S$ with the same speed and in the same direction, and the distance between $P1$ and $P2$ remains constant in $S$. We could think of $P1$ and $P2$ as the ends of a rod moving in $S$, and that too, \emph{rigidly}, since its length remains constant in $S$. 

What are the motions of $P1$ and $P2$ in $S'$? Is the distance between them constant in $S'$ too, so that the rod remains rigid in $S'$? Indeed the motions of $P1$ and $P2$ are uniform in $S'$ too, since they are given by 
\begin{eqnarray}
\label{MP1}
P1:\{( \beta (x_0 + ut - vt),0,0,\beta(t - \frac{v (x_0 + ut)}{c^2})) \}
\end{eqnarray}
and 
\begin{eqnarray}
\label{MP2}
P2:\{( \beta (x_0 + l + u\bar{t} - v\bar{t}),0,0,\beta(\bar{t} - \frac{v (x_0 + l +  u\bar{t})}{c^2})) \}.
\end{eqnarray}
Their common speed is given by 
\begin{eqnarray}
\label{SP}
\frac{u - v}{(1 - \frac{uv}{c^2})}.
\end{eqnarray}
The $S'$- distance between $P1$ and $P2$ at a time $t'$ in $S'$ is given by 
\begin{eqnarray}
\label{DIST}
\beta (x_0 + l + u \bar{t} - v \bar{t}) - \beta (x_0 + ut - vt)
\end{eqnarray}
where $t$ and $\bar{t}$ are related to $t'$ by 
\begin{eqnarray}
\label{TIME1}
t' = \beta ( t - \frac{v (x_0 + ut)}{c^2}) = \beta ( \bar{t} - \frac{v (x_0 + l + u\bar{t})}{c^2}).
\end{eqnarray}
The distance calculates out to be
\begin{eqnarray}
\label{DIST1}
\frac{1}{\beta(1 - \frac{uv}{c^2})} l.
\end{eqnarray}
Thus, the rod is observed to stay rigid in $S'$ too. But is its length in $S'$ necessarily smaller than its length $l$ observed in $S$? Denoting the factor multiplying $l$ in (\ref{DIST1}) by $k(u)$, the function $k$ has the following values:
\begin{eqnarray}
k(\frac{c^2}{v}) &= \infty, \\
k(c) &=  \frac{1}{\beta(1-v/c)} = \sqrt{\frac{1+v/c}{1-v/c}} > 1, \\
k(v) &=  \frac{1}{\beta(1-v^2/c^2)} = \beta > 1, \\
k(0) &=  \frac{1}{\beta} < 1, \\
k(-c) &=  \frac{1}{\beta(1+v/c)} = \sqrt{\frac{1-v/c}{1+v/c}} < 1, \\
k(-\infty) &= 0,
\end{eqnarray}
with 
\begin{eqnarray}
k(-c) < k(0) < 1 < k(v) < k(c).
\end{eqnarray}

The case $u=0$ corresponds to the rod being at rest in $S$ and its length in $S'$ is observed to be smaller than its length in $S$, but if $u=v$, the rod is at rest in $S'$, its length in $S'$ is observed to be \emph{larger} than its length in $S$. Further, there is a particular value of $u$, namely: 
\begin{eqnarray}
\bar{u} = \frac{c^2}{v}\left [  1 - \sqrt{1-v^2/c^2} \right],
\end{eqnarray}
such that $k(\bar{u}) = 1$, and $0<\bar{u}<v$. Thus, there is a speed $\bar{u}$ for which the rod is observed to be moving in \emph{both} $S$ and $S'$, but its length is observed to be the \emph{same} in both.

So, we can have not only a \emph{contraction} but also an \emph{expansion} and even \emph{invariance}!

\section{Length of a rod changes and does not change!}
It will be seen after a little thought that in the deductions above, it is not necessary that the points $P1$ and $P2$, or the rod whose ends they might be, be in uniform motion (or at rest) for \emph{all} time $t$. It is enough if there is uniform motion (or rest) over a sufficiently long time-interval or duration. Therefore, one can imagine a motion of rod which is rigid in $S$, i.e., remains constant in length, say, $l$, all the time; it ``starts'' in a state of rest, say, then smoothly accelerates to a state of uniform motion at the appropriate velocity $\bar{u}$, stays in that state for a while, and then smoothly accelerates again to a state of uniform motion with speed $v$, so that it is finally at rest in $S'$.  As seen from $S'$, the length of the rod will start with a value that is less than $l$, changing smoothly to $l$ after some time, and then changing smoothly to a value greater than $l$ finally. Thus, the rod will be seen to change its length in $S'$ as it moves remaining rigid in $S$. However, in between the spells of acceleration indicated above, its length in $S'$ will remain constant.  The ``obsever'' of the system $S'$ could then ascribe this change in length to acceleration of the moving rod. However, the same observer \emph{could} also calculate the length of the rod as it would be observed by yet another observer of another system $S''$, and \emph{could} in particular calculate the velocity $-v$ of the system $S''$ relative to $S'$ such that in $S''$ the rod would have a constant length in spite of its acceleration.

We may conclude, therefore, that the ``change'' in length is purely a \emph{kinematical} fact, arising out of the manner in which the two systems $S$ and $S'$ and their coordinates and times are related, and we need not look for any \emph{dynamical} reason for the change in either system.

\section{What about non-uniform motion of the rod?}
We are thus led to ask the following question:  is it possible for two points $P1$ and $P2$ (or a rod) to have \emph{continuously} accelerated motion in $S$ while maintaining a constant distance between them in $S$, and yet to be seen to maintain a constant distance between them in \emph{$S'$ also}? We explore the special situation when the motion in $S$ is along the $x$- axis (and so, in $S'$ the motion is along the $x'$- axis), and so, we suppress the $y$- and $z$- coordinates in the calculations below.

Let the motions of two points $P1$ and $P2$ in $S$ be given by two functions $x_1(t)$ and $x_2(t)$ with $x_2(t)=x_1(t) + l$ for all $t$, so that the distance between them remains contant in $S$. Let the two functions $x_1'(t')$ and $x_2'(t')$ describe their motions in $S'$. We seek conditions under which the difference $x_2'(t')-x_1'(t')$ will be constant. Now, let $t_1$ and $t_2$ denote the times in $S$, corresponding to a common time $t'$ in $S'$, for the two motions, so that we have, from (\ref{ILT}): 
\begin{eqnarray}
\label{M1}
\eqalign
t_1 &= \beta (t' + \frac{vx_1'(t')}{c^2})\ ,\quad  x_1(t_1) = \beta(x_1'(t') + vt') \\
\label{M2}
t_2 &= \beta (t' + \frac{vx_2'(t')}{c^2})\ ,\quad  x_2(t_2) = \beta(x_2'(t') + vt') 
\end{eqnarray}
and so, 
\begin{eqnarray}
\label{M3}
t_2 - t_1 = \frac{\beta v}{c^2} (x_2'(t') -x_1'(t') ). 
\end{eqnarray}
Thus, the distance $x_2'(t')-x_1'(t')$ between the two points in $S'$ will be constant, say, $l'$, if and only if $(t_2 - t_1)$ is constant, say $\alpha$, where $\alpha = \beta v l'/c^2$. But then 
\begin{eqnarray}
x_2(t_2) - x_1(t_1) &= \beta (x_2'(t') - x_1'(t')) \\
&= \frac{c^2 \alpha}{v}.
\end{eqnarray}
Since $t_2 - t_1 = \alpha$, we can write the above as 
\begin{eqnarray}
x_2(t_1 + \alpha) - x_1(t_1) = \frac{c^2 \alpha}{v}, 
\end{eqnarray}
and since
\begin{eqnarray}
x_2(t_1 + \alpha) = x_1(t_1 + \alpha) + l,
\end{eqnarray}
finally
\begin{eqnarray}
\label{M4}
x_1(t_1 + \alpha)  - x_1(t_1) =  \frac{c^2 \alpha}{v} - l.
\end{eqnarray}
Now (\ref{M4}) must hold for all times $t_1$. (To see why, imagine choosing first the time instant $t_1$, then using (\ref{LT}), determining the corresponding $t'$ in $S'$, and finally determining the corresponding $t_2$ in $S$ using (\ref{ILT})). Thus, we have proved the following necessary condition.

\begin{prop}
If the motions $x_1(t)$ of $P1$ and $x_2(t)=x_1(t)+l$ of $P2$ are such that the distance between them in $S'$ is a constant, say, $l'$, then with $\alpha = \frac{\beta v l'}{c^2}$, we must have, for all $t_1$: 
\begin{eqnarray}
\label{M5}
x_1(t_1 + \alpha)  - x_1(t_1) =  \frac{c^2 \alpha}{v} - l.
\end{eqnarray}
\end{prop}

If the function $x_1$ is continuously differentiable, from (\ref{M5}), we see immediately that the derivative of $x_1$, i.e., the \emph{velocity} , of $P1$, is a periodic function of the time in $S$, with period $\alpha$. 

In the special case when $x_1$ is a uniform motion with constant speed $u$, $x_1(t_1) = ut_1$ we have 
\begin{eqnarray}
u \alpha  =  \frac{c^2 \alpha}{v} - l,
\end{eqnarray}
and so 
\begin{eqnarray}
l' = \frac{1}{\beta(1 - \frac{uv}{c^2})} l,
\end{eqnarray}
which agrees with (\ref{DIST1}). (Of course, a constant function is periodic with any arbitrary period). 

Note that we can no longer talk about the entire rod remaining rigid, i.e., all the points of the rod including its end points maintaining a constant distance between each other in $S'$, because (\ref{M5}) cannot hold for all points. So, we can only talk about a pair of points.

Conversely, if condition (\ref{M5}) is satisfied, then with $t_2$ defined by $t_2-t_1 = \alpha$, the time instants $t_1'$ and $t_2'$ corresponding to $t_1$ and $t_2$ respectively are equal. Indeed,
\begin{eqnarray}
t_1' &= \beta (t_1 - \frac{vx_1(t_1)}{c^2}), \\
t_2' &= \beta (t_2 - \frac{vx_2(t_1 + \alpha)}{c^2}), 
\end{eqnarray}
and so
\begin{eqnarray}
x_2(t_1+\alpha) &= x_1(t_1 + \alpha + l) \\
&= [ x_1(t_1) + \frac{c^2\alpha}{v} - l  ] + l \\
&= x_1(t_1) + \frac{c^2\alpha}{v},
\end{eqnarray}
so,
\begin{eqnarray}
t_2'-t_1' &= \beta [(t_2 - t_1 ) - \frac{v}{c^2} \frac{c^2\alpha}{v} ] \\
&= \beta (\alpha - \alpha )\\
&= 0.
\end{eqnarray}
Thus, we have proved the following proposition.

\begin{prop}
The motions $x_1(t)$ and $x_2(t)$ of two points with a constant distance $l=x_2(t) - x_1(t) $ between them in $S$ are transformed into motions with a constant distance $l'=x_2'(t') - x_1'(t')$  between them in $S'$ given by $l' = {c^2\alpha}/{\beta v}$, if and only if the motion $x_1(t)$ (with a similar relation for $x_2(t)$) satisfies for some $\alpha> 0$, 
\begin{eqnarray}
\label{M6}
x_1(t+ \alpha) - x_1(t) &=  \frac{c^2\alpha}{v} - l,
\end{eqnarray}
and thus has a speed which is periodic with period $\alpha$. Further, as we can expect, with 
\begin{eqnarray}
\alpha'& = \frac{\beta v l}{c^2}, 
\end{eqnarray}
we have
\begin{eqnarray}
x_1'(t' + \alpha') - x_1'(t') &= \frac{c^2\alpha'}{v} - l',  
\end{eqnarray}
and so the motion $x_1'(t')$ has a periodic speed with period $\alpha'$ in $S'$, and, symmetrically, 
\begin{eqnarray}
l &= \frac{c^2 \alpha'}{\beta v}.
\end{eqnarray}
\end{prop}
Note that when the motion $x_1(t)$ is uniform, the speed is constant and there is no period of the function. However, (\ref{M6}) still holds. 
\begin{corollary}
If, in particular, $\alpha = \frac{\beta v l }{c^2}$ and $x_1(t+\alpha)-x_1(t)=(\beta - 1)l$, then $l'=l$, so that there is no change in length from $S$ to $S'$.
\end{corollary}
\begin{corollary}
If, in particular, $x_1(t)$ itself is periodic, then from (\ref{M6}), we get $\frac{c^2\alpha}{v}= l$, and thus, if in addition the period is given by $\alpha = \frac{lv}{c^2}$, then $l'=\frac{1}{\beta}l$ and $x_1'(t')$ is also periodic with period $\alpha'$ given by $\alpha' = \frac{l'v}{c^2} = \frac{1}{\beta}\alpha$.
\end{corollary}

\section{Is the Lorentz contraction inevitable, or, was Einstein right?}
The transformation equations (\ref{LT}) were obtained by Einstein in his pioneering paper of 1905\cite{Einstein1905}. In that paper, as a prelude to (\ref{LT}), he derived the equations
\begin{equation}
\label{ELT}
\eqalign
x' = \phi(v)\beta (x - vt), \\
y' = \phi(v)y, \\
z' = \phi(v)z, \\
t' = \phi(v)\beta (t-\frac{vx}{c^2})
\end{equation}
where $\beta=1/\sqrt{1-v^2/c^2}$. (In \cite{Einstein1905}, he used the Greek letters $\xi$, $\eta$, $\zeta$, $\tau$ instead of $x'$, $y'$, $z'$, $t'$ in (\ref{ELT}) above, although in his next expository essay \cite{Einstein1907} on the `Principle of relativity' he dropped the Greek letters and switched to the Roman letters. With this change of letters, we quote him from \cite{Einstein1905} below.)

There are some significant differences between these two writings of Einstein, even as regards the `Kinematical Part' of the two. In \cite{Einstein1905}, `I. Kinematical Part' has 5 sections: \S 1. Definition of Simultaneity \S 2. On the Relativity of Lengths and Times \S 3. Theory of the Transformation of Co-ordinates and Times from a Stationary System to another System in Uniform Motion of Translation Relatively to the Former \S 4. Physical Meaning of the Equations Obtained in Respect to Moving Rigid Bodies and Moving Clocks \S 5.  The Composition of Velocities. In \cite{Einstein1907}, `I. Kinematical Part' has 5 sections: \S 1. The principle of the constancy of the velocity of light. Definition of time. The principle of relativity \S 2. General remarks concerning space and time \S 3. Transformation of coordinates and time \S 4. Consequences from the transformation equations that concern rigid bodies and clocks \S 5. Addition theorem of velocities \S 6. Applications of the transformation equations to some optical problems. To aid comparison of the two texts, we quote below at length from the translations of the two papers in\cite{PerrettJeffery1923} and \cite{Schwartz1977}, respectively.
\\ \\
\noindent In \S 1 of \cite{Einstein1905}, we have the admonition: 
\begin{quote}
Now we  must bear carefully in mind that a mathematical description of this kind has no physical meaning unless we are quite clear as to what we understand by ``time.''
\end{quote}

\noindent He then goes on to define the synchronization of two clocks at two places $A$ and $B$: 
\begin{quote}
We have not defined a common ``time'' for A and B, for the latter cannot be defined at all unless we establish {\em by definition} that the ``time'' required by light to travel from A to B equals the ``time'' it requires to travel from B to A\@.  Let a ray of light start at the ``A time'' $t_{\rm A}$ from A towards B, let it at the ``B time'' $t_{\rm B}$ be reflected at B in the direction of A, and arrive again at A at the ``A time'' $t'_{\rm A}$.

In accordance with definition the two clocks synchronize
if
\[
t_{\rm B}-t_{\rm A}=t'_{\rm A}-t_{\rm B}.
\]
\end{quote}

\noindent In \S 1 of \cite{Einstein1907}, on the other hand, he says: 
\begin{quote}
We consider clocks, at rest, relative to the coordinate system, arranged at many points. These are all to be equivalent, i.e., the difference of the readings of any two such clocks are to remain unaltered, when they are arranged near each other. If we imagine these clocks stationed in any manner, then provided they are arranged with sufficiently small separations, the ensemble of clocks allows the temporal  labelling of an arbitrary point-event -- namely, by means of the adjacent clocks.

The sum total of these clocks-readings does not, however, provide us as yet with a ``time'', as it is needed for the purposes of Physics. We require in addition a rule according to which these clocks are to be set with respect to each other. 

We assume now that \emph{the clocks can be so regulated, that the propagation velocity in empty space of every light ray- when measured with these clocks- is everywhere equal to a universal constant $c$}, provided the coordinate system is not accelerated. If $A$ and $B$ are two points occupied by clocks at rest in the coordinate system and at a distance $r$ apart, and if $t_A$ is the reading of the clock at $A$ when a light ray propagating through a vacuum in the direction $AB$ reaches the point $A$, and $t_B$ is the reading of the clock at $B$ when the light ray arrives at $B$, then regardless of the state of motion of the light source or of other bodies, one always has 
\begin{eqnarray}
r/(t_B-t_A) = c.
\end{eqnarray}

That the assumption just made, which we shall call the ``principle of the constancy of the velocity of light'', is actually satisfied in nature, is not at all self-evident, but it is made probable - at least for a coordinate system in a definite state of motion - by the experimental confirmations of Lorentz's theory \cite{Lorentz1895}, which is based on the assumption of an absolutely stationary ether.
\end{quote}

\noindent It seems to us that this difference may be the result of (a) Einstein's becoming familiar with Lorentz 1904 \cite{Lorentz1904}, but also perhaps (b) his realization that contemporary physicists, who were more interested in the ``theory of the electron'', were unlikely to pay heed to his ``admonition'' and express desire for ``physical'' meaning. (We note that in \cite{Einstein1907}, Einstein does not mention Poincare although \cite{Lorentz1904} does so.) This difference in attitude perhaps resulted in differences in the third section of the two papers. The derivation of the transformation equations in \cite{Einstein1905} involved physically meaningful Gedanken experiments of light rays getting emitted, being reflected, and arriving back at the starting place. Instead in \cite{Einstein1907}, we have:
\begin{quote}
We now conclude immediately from our knowledge of the position of the coordinate planes of $S'$ relatively to $S$, that every pair of the following set of equations is equivalent: 
\begin{eqnarray}
x' = 0 \quad \textrm{ and } \quad &x-vt = 0; \\
y' = 0 \quad \textrm{ and } &y = 0;\\
z' = 0 \quad \textrm{ and } &z = 0.
\end{eqnarray}
Hence three of the sought transformation equations are of the form:
\begin{eqnarray}
x' &= a(x-vt),\\
y' &=by,\\
z' &=c\textrm{ (\emph{sic}) }z.
\end{eqnarray}
Since the velocity of propagation of light in empty space equals $c$ with respect to both reference systems, the two equations 
\begin{eqnarray}
x^2+y^2+z^2 = c^2t^2
\end{eqnarray}
and
\begin{eqnarray}
x'^2+y'^2+z'^2 = c^2t'^2
\end{eqnarray}
must be equivalent. From this and from the above-found expressions for $x'$, $y'$, $z'$ one concludes after simple calculations, that the sought transformation equations must be of the form:
\begin{eqnarray}
t'&=\phi(v)\beta[t-(vx/c^2)], \\
x'&=\phi(v)\beta(x-vt),\\
y'&=\phi(v)y,\\
z'&=\phi(v)z,
\end{eqnarray}
where we have set
\begin{eqnarray}
\beta = [1-(v/c)^2]^{-1/2}.
\end{eqnarray}
\end{quote}
In \cite{Einstein1905}, he \emph{derives} $\xi^2 + \eta^2 + \zeta^2 = c^2\tau^2$ from $x^2+y^2+z^2 = c^2t^2$ using the transformation equations, whereas in \cite{Einstein1907}, he \emph{derives} the transformation equations from the requirement of $x^2+y^2+z^2 = c^2t^2$ and $x'^2 + y'^2 + z'^2 = c^2t'^2$ being equivalent. 

More significantly for our present discussion of ``contraction'', what follows from the  transformation equations is almost identical in \cite{Einstein1905} and \cite{Einstein1907}. 
\\ \\
\noindent In \cite{Einstein1905} we have:
\begin{quote}
In the equations of transformation which have been developed there
enters an unknown function $\phi$ of $v$, which we will now determine.

For this purpose we introduce a third system of co-ordinates
\pr{K}, which relatively to the system $k$ is in a state of parallel
translatory motion parallel to the axis of $\Xi$, 
such that the origin of
co-ordinates of system \pr{K} moves with velocity $-v$ on the axis of $\Xi$.  At the time $t=0$ let all three origins coincide, and when $t=x=y=z=0$
let the time $t'$ of the system \pr{K} be zero.  We call the
co-ordinates, measured in the system \pr{K}, $x'$, $y'$, $z'$, and by a twofold
application of our equations of transformation we obtain
\[
\begin{array}{lllll}
t' & = & \phi(-v)\beta(-v)(\tau+v\xi/c^2) & = &\phi(v)\phi(-v)t,\\
x' & = & \phi(-v)\beta(-v)(\xi+v\tau) & = & \phi(v)\phi(-v)x,\\
y' & = & \phi(-v)\eta & = & \phi(v)\phi(-v)y,\\
z' & = & \phi(-v)\zeta & = & \phi(v)\phi(-v)z.\\
\end{array}
\]

Since the relations between $x'$, $y'$, $z'$ and $x$, $y$, $z$
do not contain the time $t$, the systems K and \pr{K} are at rest
with respect to one another, and it is clear that the
transformation from K to \pr{K} must be the identical
transformation.  Thus

\[
\phi(v)\phi(-v)=1.
\]

\noindent
We now inquire into the signification of $\phi(v)$.  We give our attention
to that part of the axis of Y of system $k$ which lies between
$\xi=0, \eta=0, \zeta=0$ and $\xi=0, \eta=l, \zeta=0$.
This part of the axis of Y is a
rod moving perpendicularly to its axis with velocity $v$ relatively to
system K\@.  Its ends possess in K the co-ordinates

\[
x_1=vt,\ y_1=\frac{l}{\phi(v)},\ z_1=0
\]
and
\[
x_2=vt,\ y_2=0,\ z_2=0.
\]

\noindent
The length of the rod measured in K is therefore $l/\phi(v)$; and this
gives us the meaning of the function $\phi(v)$.  From reasons of symmetry
it is now evident that the length of a given rod moving
perpendicularly to its axis, measured in the stationary system, must
depend only on the velocity and not on the direction and the sense of
the motion.  The length of the moving rod measured in the stationary
system does not change, therefore, if $v$ and $-v$ are interchanged.
Hence it follows that $l/\phi(v)=l/\phi(-v)$, or

\[
\phi(v)=\phi(-v).
\]
\noindent
It follows from this relation and the one previously found that
$\phi(v)=1$,
so that the transformation equations which have been found become
\begin{equation}
\eqalign
\tau = \beta(t-vx/c^2), \\
\xi  = \beta(x - vt), \\
\eta = y, \\
\zeta = z, 
\end{equation}
\noindent
where
\[
\beta=1/\sqrt{1-v^2/c^2}.
\]
\end{quote}
\noindent In \cite{Einstein1907} we have (with two surprising footnotes) the following. (Note that $K$, $k$, and $K'$ of \cite{Einstein1905} have become $S$, $S'$, and $S''$ in \cite{Einstein1907}.)
\begin{quote}
We shall now determine the function of $v$ which still remains undetermined. If we introduce a third reference system $S''$, that is equivalent to $S$ and $S'$, moves with the velocity $-v$ relative to $S'$ and is oriented relative to $S'$ as $S'$ is to $S$, we obtain by a double application of the equations arrived at above
\begin{eqnarray}
t'' &= \phi(v)\phi(-v)t, \\
x'' &= \phi(v)\phi(-v)x, \\
y'' &= \phi(v)\phi(-v)y, \\
z'' &= \phi(v)\phi(-v)z.
\end{eqnarray}
Since the origins of coordinates of $S$ and $S''$ remain in coincidence, and since the axes have the same orientation and the systems are ``equivalent'', therefore this transformation is the identity [Einstein's footnote: This conclusion is based on the physical assumption that the length of a measuring rod as well as the rate of a clock do not suffer any lasting change by being set in motion and then brought back to rest.], so that 
\begin{eqnarray}
\phi(v)\phi(-v) = 1.
\end{eqnarray}
Since, moreover, the relationship between $y$ and $y'$ cannot depend on the sign of $v$, 
\begin{eqnarray}
\phi(v)&=\phi(-v).
\end{eqnarray}
Therefore [Einstein's footnote: Obviously $\phi(v)=-1$ does not enter into consideration.], $\phi(v)=1$, and the transformation equations read
\begin{equation}
\eqalign
t'  =  \beta(t-vx/c^2), \\
x'   =  \beta(x - vt), \\
y'  =  y, \\
z'  =  z, 
\end{equation}
\noindent
where
\[
\beta=[1-(v/c)^2]^{-1/2}.
\]
\end{quote}
\noindent Thus, in both \cite{Einstein1905} and \cite{Einstein1907}, he has looked at a ``rod'' lying \emph{transverse} to the direction of relative motion of the two systems, and \emph{assumed} that there is no change in the length of the rod from one system to the other. In neither does he talk about ``contraction'' in the direction of motion. In \cite{Einstein1905}, he \emph{does not} talk of a rod lying at rest on the $x'$ axis of $S'$; instead he envisages `a rigid sphere of radius $R$, at rest relatively to the moving system $k$' and concludes that the `sphere,  therefore has in a state of motion - viewed from the stationary system - the form of an ellipsoid of revolution with the axes
\begin{eqnarray}
R\sqrt{(1-v^2/c^2)},\ R,\ R.'
\end{eqnarray}
He does say however, that `Thus, whereas the $Y$ and $Z$ dimensions of the sphere (and therefore of every rigid body of no matter what form) do not appear modified by the motion, the $X$ dimension appears shortened in the ratio $1:\sqrt{(1-v^2/c^2)}$, i.e., the greater the value of $v$, the greater the shortening.'
\\ \\
\noindent In \cite{Einstein1907}, on the other hand, he starts off his \S 4 with:
\begin{quote}
Consider a body at rest relative to $S'$. Let $x_1'$, $y_1'$, $z_1'$ and $x_2'$, $y_2'$, $z_2'$ be the coordinates of two of its material points referred to $S'$. Between the coordinates $x_1$, $y_1$, $z_1$ and $x_2$, $y_2$, $z_2$ of these points relative to $S$,  there obtain at each time $t$ of $S$, according to the above-derived transformation equations, the relations 
\begin{eqnarray}
x_2-x_1 &= [1-(v/c)^2]^{-1/2}(x_2'-x_1'),\\        
y_2-y_1 &= y_2'-y_1',\\
z_2-z_1 &= z_2'-z_1'.
\end{eqnarray}

The kinematic shape of a body considered to be in a state of uniform translation depends thus on its velocity relative to the reference system; namely, by differing from its geometric shape in being contracted in the direction of the relative motion in the ratio $1:\sqrt{(1-v^2/c^2)}$.
\end{quote}

\noindent In both \cite{Einstein1905} and \cite{Einstein1907}, he goes on to talk about `slowing of the clock', but once again, there is a difference. In \cite{Einstein1905}, he considers a clock completing a round trip, whereas in \cite{Einstein1907}, he considers only the two clocks at the origins of the two systems. 

To get back to our discussion of ``contraction'', we suggest that there is a flaw in Einstein's argument regarding the `the unknown function $\phi$ of $v$' in both \cite{Einstein1905} and \cite{Einstein1907}. His statement that `it is clear that the transformation from $K$ to $K'$ must be the identical transformation' is only an \emph{assumption}, and not \emph{proved} on the basis of the transformation equations. The coordinates and time in $K'$ could well be only \emph{scaled versions} of the coordinates and time in $K$, \emph{both} coordinates and time being scaled by the \emph{same} scale factor. Why, then, did he say `it is clear'? We suspect it is because of the manner in which he has started thinking about the two systems $K$ and $k$. Thus, in \S 3 of \cite{Einstein1905}, he starts off with:

\begin{quote}
Let us in ``stationary'' space take two systems of co-ordinates,
i.e.\ two systems, each of three rigid material lines,
perpendicular to one another, and issuing from a point.  Let the
axes of X of the two systems coincide, and their axes of Y and Z
respectively be parallel.  Let each system be provided with a
rigid measuring-rod and a number of clocks, and let the two
measuring-rods, and likewise all the clocks of the two systems,
be in all respects alike.

Now to the origin of one of the two systems ($k$) let a constant
velocity $v$ be imparted in the direction of the increasing $x$ of the
other stationary system (K), and let this velocity be communicated to
the axes of the co-ordinates, the relevant measuring-rod, and the
clocks.  To any time of the stationary system K there then will
correspond a definite position of the axes of the moving system, and
from reasons of symmetry we are entitled to assume that the motion of
$k$ may be such that the axes of the moving system are at the time
$t$
(this ``$t$'' always denotes a time of the stationary system) parallel to
the axes of the stationary system.
\end{quote}

\noindent He then lets `at the time $t=0$ all the three origins coincide'. If we were to consider $K$ and $k$ (and $K'$) to be somehow ``given'' already, then with $\phi(v)=1$, we will have no `change' in the $y$- and $z$- coordinates from one system to the other; this goes well with the visualization that the origin $O'$ moves in the direction of the positive $x$- axis of $K$, the $y'$- and $z'$- axes of $k$ remaining parallel to the $y$- and $z$- axes of $K$. But, then we should be \emph{disturbed} about the $x$- coordinate not remaining unchanged from one system to the other (of course, the $x$- coordinate could change in the Galilean fashion). On the other hand, if we choose $\phi(v)=1/\beta$, then the $y$- and $z$- coordinates will change,  the $x$- coordinate will change in Galilean fashion, and the time will not change. Indeed, with this choice of $\phi(v)$, the reading of the moving clock at $O'$ will agree with the reading of the $K$- stationary clocks of $K$ that coincide with $O'$ at each moment, and in this sense, we could have a ``universal clock''. Note, however, that the time of an \emph{event}, in general, is not universal since we will still have 
\begin{eqnarray}
\tau = t - \frac{vx}{c^2}, 
\end{eqnarray}
so the $k$- time of an event will depend not only on the $K$- time of the event but also on its $x$- coordinate in $K$. The $x'$- coordinate in $K$ will depend on the $x$- coordinate and time of the event in $K$, in the Galilean fashion:
\begin{eqnarray}
\xi = x - vt.
\end{eqnarray}
The $y$- and $z$- coordinates will undergo a change:
\begin{eqnarray}
\eta &= \frac{1}{\beta} y , \\
\zeta &= \frac{1}{\beta} z.
\end{eqnarray}
Thus, for a given system $K$, from among \emph{all} systems $k$ which move with velocity $v$ in the $x$- direction, we can ``choose'' one for which the \emph{new} transformation equations above will hold. Selecting for each $v$ an appropriate moving system, the highly desired ``group property'' of the transformations will hold for the ensemble of such systems.

But, we will argue that even if we accept, with Einstein, that 
\begin{eqnarray}
\phi(v) \phi(-v) = 1,
\end{eqnarray}
it does \emph{not} follow that $\phi(v) = \phi(-v)$. Indeed, Einstein has concluded that if $l$ is the length of a rod lying along the $y$ axis of $k$ then its length measured in $K$ is $l/\phi(v)$, $v$ being the velocity of $k$ relative to $K$. Thus, 
\begin{eqnarray}
\frac{\textrm{length of rod in } k}{\textrm{length of rod in } K} = \phi(v), \ v \textrm{ is velocity of } k \textrm{ relative to } K.
\end{eqnarray}
But then $K'$ moves with velocity with $-v$ relative to $k$. So 
\begin{eqnarray}
\frac{\textrm{length of rod in } K'}{\textrm{length of rod in } k} = \phi(-v), \ -v \textrm{ is velocity of } K' \textrm{ relative to } k.
\end{eqnarray}
``Multiplying'' the relations above we get
\begin{eqnarray}
\frac{\textrm{length of rod in } K'}{\textrm{length of rod in } K} = \phi(v)\phi(-v) = 1, 
\end{eqnarray}
as expected!

Thus, Einstein's appeal to ``symmetry'' is only an \emph{assumption}, and we could therefore,  live with new transformation equations
\begin{eqnarray}
\label{NLT}
\eqalign
x' &=  x - vt \\
y' &= \frac{1}{\beta} y \\
z' &= \frac{1}{\beta} z \\
t' &= t-\frac{vx}{c^2},
\end{eqnarray}
and have, as a consequence, ``contraction'' or ``expansion'' in the $y$- and $z$- directions, but no ``contraction'' or ``expansion'' in the $x$- direction, and most importantly, no \emph{slowing down} or \emph{speeding up} of ``clocks''. Lorentz \cite{Lorentz1904} had a factor `$l$' like Einstein's $\phi(v)$, and concluded that $l=1$ for \emph{dynamical} reasons. But Lorentz's mechanics was Newtonian mechanics, leading to ``longitudinal'' and ``transverse'' masses. With Planck's new definition of force, or rather, a new Second Law of Motion, namely 
\begin{equation}
\mathbf{F} = \frac{d}{dt}(\frac{m_0\mathbf{v}}{\sqrt{1-v^2/c^2}}),
\end{equation}
in fact, this $l$ will have to be $1/\beta$. This Planckian ``scaling'' of mass can also be done for concepts like frequency of a source of light and life-time of a particle. Thus we could \emph{postulate} that there is an absolute or universal frequency $\nu_0$ associated with a particular light-source such that in a \emph{single} frame of reference, if the source moves with velocity $v$, its frequency in that frame of reference will be $\nu_0/\sqrt{1-v^2/c^2}$. Similarly, we could postulate that there is an absolute or universal life-time $\tau_0$ associated with a particular kind of particle such that its observed life-time at velocity $v$ in a single frame of reference will be $\tau_0/\sqrt{1-v^2/c^2}$. 

In conclusion, we only point out that Einstein had left out any \emph{physical} considerations involved in assigning coordinates to points of space; he seemed to be satisfied letting it rest on `the employment of rigid standards of measurement and the methods of Euclidean geometry'. We made an attempt to remedy this situation in our \cite{Agashe1} and \cite{Agashe2}.

\end{document}